\documentclass[a4paper]{article}
\usepackage{fullpage}
\usepackage{natbib}
\usepackage{graphicx}
\author{P. B. Carolan,$^{1}$\thanks{E-mail:patrick.carolan@nuigalway.ie} 
M.P. Redman, $^{1,}$$^{2,}$$^{3}$
E. Keto, $^{4}$
J.M.C. Rawlings, $^{3}$
\\
\\
\small{$^{1}$Centre for Astronomy, National University of Ireland Galway, Galway, Ireland.}\\
\small{$^{2}$School of Cosmic Physics, Dublin Institute for Advanced Studies, Merrion,Square, Dublin 2, Ireland.}\\
\small{$^{3}$Department of Physics \& Astronomy, University College London, Gower Street, London WC1E 6BT UK.}\\
\small{$^{4}$Harvard-Smithsonian Center for Astrophysics, 60 Garden Street, Cambridge MA 02138, USA.}}
\title{CO abundances in a protostellar cloud: freeze-out and desorption in the envelope and outflow of L483}
\begin{document}
\maketitle
\begin{abstract}
CO isotopes are able to probe the different components in protostellar clouds. These components, core, envelope and outflow have distinct physical conditions and sometimes more than one component contributes to the observed line profile. In this study we determine how CO isotope abundances are  altered by the physical conditions in the different components. We use a 3D molecular line transport code to simulate the emission of four CO isotopomers, $^{12}$CO $J=2\rightarrow1$, $^{13}$CO $J=2\rightarrow1$, C$^{18}$O $J=2\rightarrow1$ and C$^{17}$O $J=2\rightarrow1$ from the Class 0/1 object L483, which contains a cold quiescent core, an infalling envelope and a clear outflow. Our models replicate JCMT (James Clerk Maxwell Telescope) line observations with the inclusion of freeze-out, a density profile and infall. Our model profiles of $^{12}$CO and $^{13}$CO have a large linewidth due to a high velocity jet. These profiles replicate the process of more abundant material being susceptible to a jet. C$^{18}$O and C$^{17}$O do not display such a large linewidth as they trace denser quiescent material deep in the cloud.
\end{abstract}

%\begin{keywords}
%radiative transfer -- stars: formation -- ISM: abundances -- ISM: jets and outflows -- ISM: individual: L483 -- ISM: kinematics and dynamics
%\end{keywords}

\section{Introduction}
Molecules, particularly CO, are used as tracers of H$_2$ and thus of gas density in cold dark clouds. CO is highly abundant with a low critical density and typically exhibits strong optical depth effects in cold dark molecular clouds. The four most common CO isotopes differ in abundance by as much as three orders of magnitude. Thus they become optically thick at different column densities. Taken together, observations of CO isotopes can trace the gas density in all the main components of cold, dark clouds: the intermediate optical depth envelope, the high optical depth core, and the optically thin bipolar outflows. However, we know from observational and theoretical studies that the abundance of CO depends on conditions in the clouds such as, shock heating \citep{van_dishoeck_95,nisini_07}, U.V. excitation \citep*{goldsmith_07}, freeze-out \citep*{lee_04} and varies from place to place. Therefore we cannot use CO as an H$_2$ tracer without understanding its chemical variation. To better understand CO variations in cold dark clouds, we observed and modeled one particular cloud, Lynds 483. 
We chose L483 as a prototype for a study of CO abundances because it is a well studied nearby ($\sim 200~{\rm pc}$) molecular cloud. It contains an IRAS source 18148-0440 that is in transition between a Class 0 and Class 1 object \citep{tafalla.et.al00}. It exhibits an infalling envelope \citep{park.et.al99,park.et.al00,tafalla.et.al00} and a slow bipolar molecular outflow \citep{fuller.et.al95,buckle.et.al99} yet the core and envelope are still cold and dense \citep{ladd.et.91,fuller.et.92,fuller&wootten00}. Thus many of the physical properties and kinematic features that are present in either less or more evolved clouds are all present in L483.  Our aim is to combine these components into a single model for L483 to strongly constrain the structure and dynamics of the system and hence then to infer the CO abundance throughout the cloud. 

\begin{figure*}
\centering
\includegraphics[width=180mm]{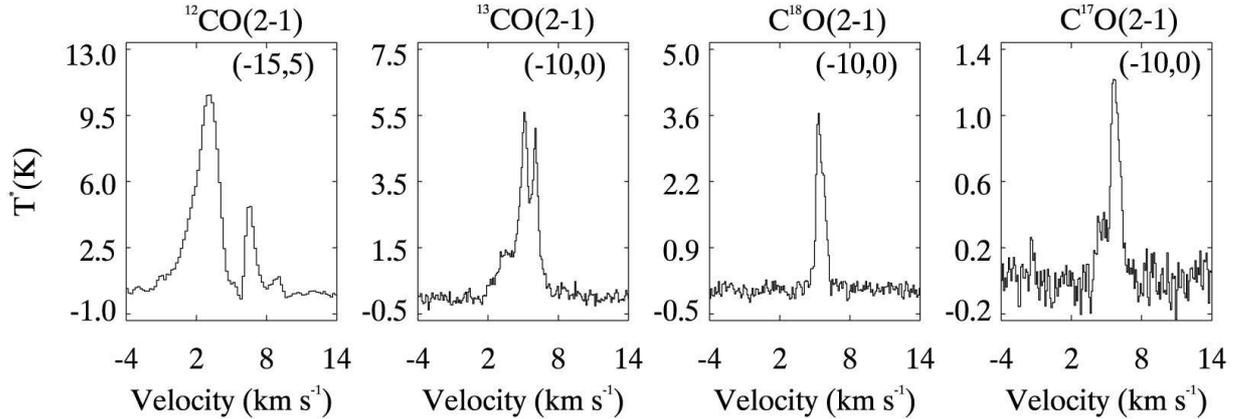}
\caption{An example line profile of each of the four isotopes used in this study where the offsets with respect to IRAS 18140-0440 are indicated in the top right corner of each panel. The $^{12}$CO y-axis is in units of $T_{r}$ and $^{13}$CO, C$^{18}$O and C$^{17}$O are in units of $T_{mb}$ }
\label{overview}
\end{figure*}

We obtained emission line profile data from L483 for transitions of the four most common CO isotopes (section ~\ref{obs}). In addition, we used an archival dust continuum emission map of L483 as an unbiased mass tracer to be compared with the CO data. We used a radiative transfer model to analyse the data and to produce synthetic spectra to be compared in detail with individual observations (section ~\ref{models}).  Estimates for density, temperature and a physical model of L483 were constrained with help from the observational literature. The abundances of the CO isotopes were then varied in the model to give a good match with the observed line profiles. This yields the abundance variation throughout L483 as well as self-consistent temperatures, densities and velocities and abundance ratios.

\section{Observations and analysis}
\label{obs}
We obtained high frequency molecular line data of three CO isotopomers in frequency switching mode. These are {${\rm ^{13}CO}~J=2\rightarrow 1$}, {${\rm C^{18}O}~J=2\rightarrow 1$} and {${\rm C^{17}O}~J=2\rightarrow 1$} in order of decreasing relative abundance.  The bulk of the observations were carried out at the JCMT, Mauna Kea, Hawaii on the nights of 2003 May 3-5. Additional data were collected in flexible observing mode. The data were reduced in the standard manner using the {\sc specx} software package. Further analysis was undertaken with the {\sc class} package of IRAM. For the majority of the observations, the system temperature was between 300 and 450 K. The maps for all the three isotopes  are centered on the position of the IRAS source, the pointing centre of which happens to be just on the blue-shifted side of the outflow. C$^{17}$O data was obtained over 40$^{\prime\prime}$ along the outflow axis and smaller data sets were obtained for C$^{18}$O and $^{13}$CO. 

\begin{table}
\centering
\caption{Species and line observed plus the number and pattern of the pointings. The cross patterns were all Nyquist sampled with separations of 10". The ${\rm ^{12}CO}$ positions are marked on Fig.~\ref{filled_contour_posn_maps}.}
\begin{tabular}{@{}lllll@{}}
\hline
Molecule & Line & Frequency & JCMT & Number of pointings\\
	& & (GHz) & receiver & and pattern\\
\hline
${\rm ^{12}CO}$ & $2\rightarrow 1$ & 230.54 & A2 & 6 individual\\
${\rm ^{13}CO}$ & $2\rightarrow 1$ & 220.40 & A3 & 5 cross\\
${\rm C^{17}O}$ & $2\rightarrow 1$ & 224.71 & A3 & 9 extended cross\\
${\rm C^{18}O}$ & $2\rightarrow 1$ & 219.56 & A3 & 5 cross\\
\hline
\end{tabular}
\label{lines}
\end{table}
\citet*{hatchell.et.al99} presented line profiles of the most common isotope {${\rm ^{12}CO}~J=2\rightarrow 1$} also obtained with the JCMT telescope. Their reduced dataset was very kindly supplied by J. Hatchell.  Table~\ref{lines} is a list of the molecular transitions and frequencies of the observations, the JCMT receiver used and the number and pattern (if any) of the pointings used to obtain spectra. The beam size was approximately 21" at the frequency of these lines. The position of the observed profiles are also centered on the IRAS source but they observe  a string of positions across both lobes of the molecular outflow. The {${\rm ^{12}CO}~J=2\rightarrow 1$} was obtained in position switching mode and some slight contamination could result from the off position used. However our results do not critically depend on this issue because any reduction of the self-absorption feature would only improve the model fit.

Finally, in addition to the molecular line data, archival SCUBA data of L483
was retrieved and reduced in the standard manner using the {\sc surf}
package in the {\sc starlink} suite of software. This data was originally published in \citet{shirley.et.al00} in the form of emission maps at $450~\mu{\rm m}$ and $850~\mu{\rm m}$.

\subsection{Overview of the line profiles}
In Fig.~\ref{overview} we present an example line profile from each of the four isotopes in order to illustrate how the different lines trace different parts of the cloud. All the lines are from the $(2 \rightarrow 1)$ transition and the full set of line profiles are presented alongside the modelling results later in the paper.

The line profile of the most abundant isotope, $^{12}$CO, exhibits very heavy self-absorption causing a complex line profile. The largest peak in the line profile is due to the molecular outflow. The outflow feature is strong despite occupying a small volume in the cloud. This is because of the different excitation and self-absorption properties in different components of the cloud plus some doppler shifting of the velocity feature away from the heavily absorbing gas close to the rest velocity of the cloud. This $^{12}$CO line is an excellent tracer of material directly influenced by the outflow. 

The $^{13}$CO profile also exhibits self-absorption due to its high optical depth. The self absorption feature is caused by overlying, low excitation temperature gas in the outer envelope that surrounds the protostellar core  \citep{myers.96,tafalla.et.al02}. We see a red and blue peak either side of the absorption with the blue wing stronger. This is usually taken to indicate infall motions are present.  Thus this line is a good tracer of the infalling and stationary parts of the outer envelope of the core. In addition, {${\rm ^{13}CO}$}  is sufficiently abundant that it can also marginally trace the outflow in the line wings.

C$^{18}$O has an intermediate optical depth and is therefore used to trace primarily regions deep in the envelope. The observed profile shows a single peak with a maximum temperature at 3K. It is abundant enough for optical depth effects to be noticeable in the shape of the line profile and a slight blue asymmetry to the profile is seen. 

{${\rm C^{17}O}$} has a low optical depth and probes the deepest into the cloud. However its abundance is correspondingly small which makes its detection harder. {${\rm C^{17}O}$} possesses a well defined hyperfine structure which is clearly seen although some mild blending is apparent. Hyperfine satellite lines are seen on either side of the strongest peak, with the right hand side component just discernible above the noise. The mild blending is due to some dynamical activity inside the envelope.

The freeze-out of molecules from the gas phase onto grain surfaces appears to be common in starless cores \citep{gibb,redman.et.al02b,bacmann.et.al02}. It is expected to occur at temperatures less than 20K \citep{sandford} and densities greater than $\sim 10^{4}~{\rm cm^{-3}}$. In order to measure the degree of CO freeze-out in the envelope of L483, we used the SCUBA dust emission data to determine the column density of H$_2$ as implied by a dust-gas ratio of 100. \citet{shirley.et.al00} used their $450~\mu{\rm m}$ and $850~\mu{\rm m}$ emission maps to produce radial column density plots by azimuthally averaging the emission data and applying a dust continuum radiative transfer model. However, to allow for a more exact comparison with our individual line profiles which lie mainly in an east-west axis across the cloud, we reanalysed the emission data and calculated column densities along a strip with the same beam width as our line data.

\begin{figure}
\centering
\includegraphics[width = 240pt]{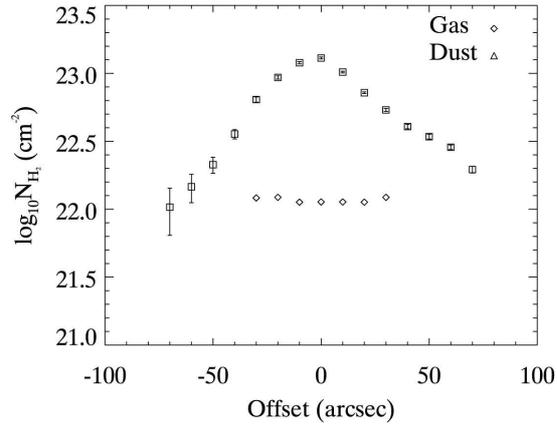}
\caption{Plot of hydrogen column density implied by SCUBA dust emission and integrated CO emission. The discrepancy between the two indicates a large degree of freeze-out of CO onto the surface of dust grains. The lack of vertical error bars is discussed in the text}
\label{freezeout}
\end{figure}

In Fig.~\ref{freezeout} we compare the column density of H$_2$ implied by the dust emission with that gas column density calculated from the best fit model. Our H$_2$ column densities derived from dust emission peak at $1.5 \times 10^{23}~{\rm cm^{-2}}$ towards the center pointing. In contrast, the column density derived from the optically thin C$^{18}$O isotope reaches only $1.0 \times 10^{22}~{\rm cm^{-2}}$ i.e. the CO is depleted by around 95\%. The gas column density profile is flat because the rate of freeze-out depends on the gas density and temperature \citep{rawlings.et.al04} and so the heaviest depletion should take place at the centre of the cloud. 

\begin{figure}
\hskip -4mm
\includegraphics[width = 245 pt]{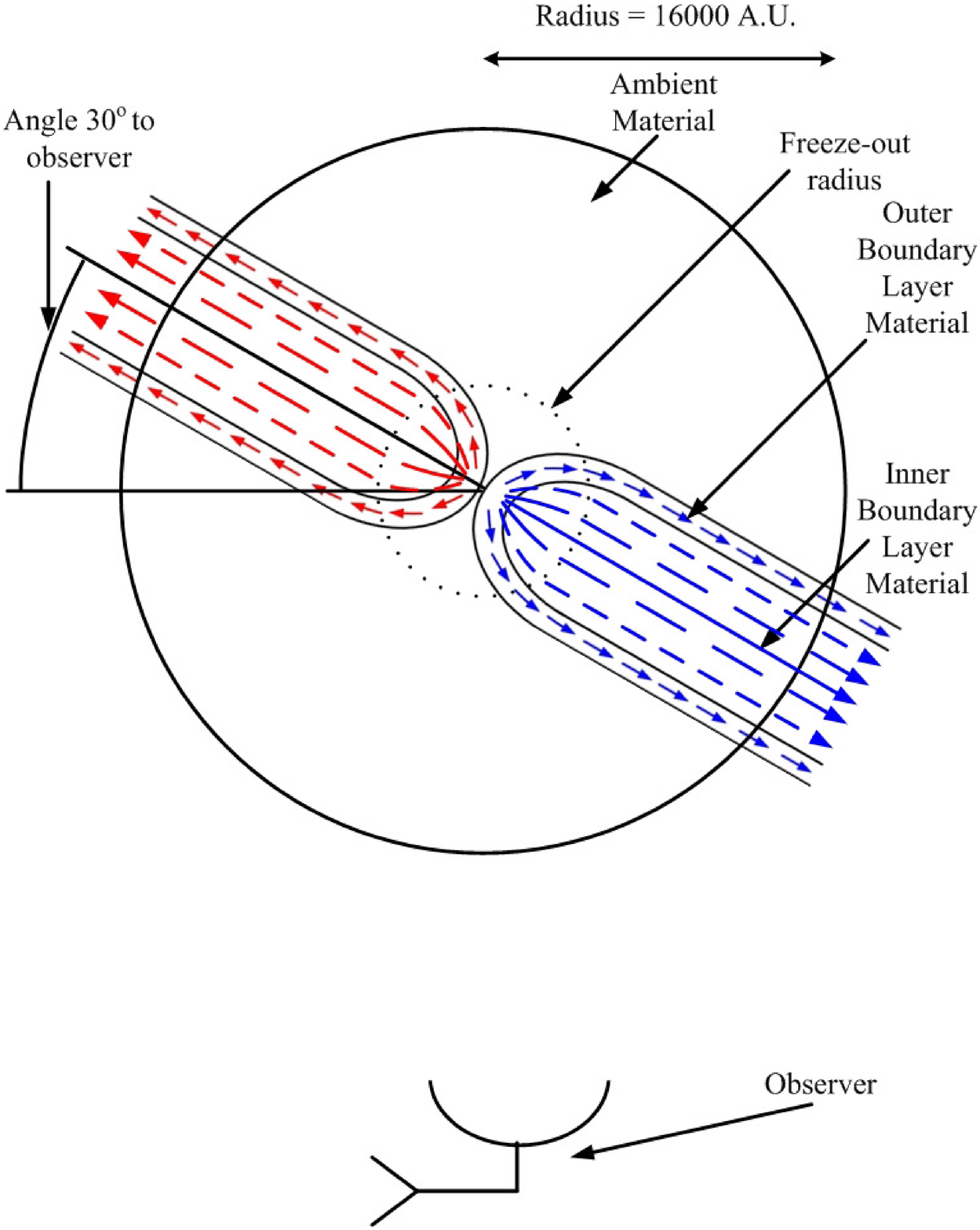}
\caption{A schematic representation of the outflow, where the radius indicated is the total cloud radius. The freeze-out radius is represented with a broken line. The lobe pointing towards us is the blue-shifted side and lies at an angle $\approx$ 30$^{\circ}$. The different length arrows represent the speed of the jet/outflow. The boundary layer is the molecular outflow as traced by our observations.}
\label{outflow_schematic}
\end{figure}

\section{Modelling}
\label{models}
In principle, there is a wealth of information encoded in the line profiles from the four different isotopes. In order to extract this information, the entire cloud needs to be modelled self-consistently. To do this requires that the geometry and physical parameters of the system are characterized as fully as possible. At each point in the model cloud the velocity, density, temperature, turbulent velocity and CO abundance are required in order to solve the radiative transfer. 

\subsection{Velocity, density and temperature structure}
{\citet{tafalla.et.al00}} argue that the evolutionary status of L483 is between a Class 0 and Class 1 object. The inner regions of the object appear to be undergoing infall.  \citet{tafalla.et.al00} observed H$_{2}$CO(2$_{21}$ $\rightarrow$ 1$_{11}$) line profiles to be self absorbed at ambient velocities with brighter blue peaks. \cite{park.et.al99} observed self-absorbed asymmetric profiles using HCN hyperfine lines to confirm this inward motion. We adopt an initial infall velocity estimate of
$0.5~{\rm km~s^{-1}}$ and central density of greater than $10^5~{\rm cm^{-3}}$ based on these results. A r$^{-2}$ density profile is used to describe the envelope density distribution since this is a simple approximation to a core containing a protostar (e.g. \citet{shirley.et.al00}).

L483 contains a molecular outflow. \citet{buckle.et.al99} presented longslit ro-vibrational spectroscopy of H$_2$ emission and discovered knotty structure along the jet. The speed of the jet was calculated to be 40 $\rightarrow$ 45 km s$^{-1}$. The H$_{2}$ emission is strongest in the blue-shifted side of the outflow, which is positioned at an angle 25$^{\circ}$ $\rightarrow$ 30$^{\circ}$ towards the observer. \citet{fuller.et.al95} mapped the outflow in $^{12}$CO (3 $\rightarrow$ 2) and in 2.12 $\mu$m molecular hydrogen emission. Again a compact collimated bipolar structure was observed along with a young driving source. VLA 6cm continuum observations were carried out  by \citet{beltran.et.al01} who found a central source with a spectral index consistent with thermal free-free emission from thermal radio jets. 

\citet{buckle.et.al99} concluded that there is a factor of at least 10 in the difference in density between the jet and the surrounding medium at their point of measurement. Since the head of the jet is well outside the dense core we adopt an under-density of between 10-100 in the molecular outflow compared with the central density of the core. Any CO emission seen in the low J observations presented here will be from the edges of the interaction region between the jet and the envelope. In molecular outflows the hottest, highest speed material is found closest to the jet axis (e.g. HH211). Two models for the origin of molecular outflows are turbulent entrainment in a boundary layer behind a leading jet-driven bow shock or a wide angle wind from the protostar. Following \citet{hatchell.et.al99}, we favour the former scenario though our analysis is not dependent on the model used.

Since there will be a velocity and temperature gradient in the outflowing CO we characterize the boundary layer region with two components: a warm low density inner boundary layer and cooler denser outer boundary layer with a linear fall-off between them. The jet is under-dense with a modest leading shock velocity ($40 - 45~{\rm km~s^{-1}}$  \citealt{buckle.et.al99}) so from momentum conservation the mixing of the jet and the denser envelope will lead to a slow molecular outflow. This is seen clearly in Fig.~\ref{overview} where the emission peak due to the outflowing CO is at very modest velocity. Thus we adopt an initial velocity estimate of only  $\sim 5~{\rm km~s^{-1}}$. The temperature is adopted to be 10K in the envelope and between 35-100K in the boundary layer. \citet{jorgensen.et.al04} used a Monte Carlo modeling technique for the envelope of L483. They found a turbulent width between $0.5 - 1.0~{\rm km~s^{-1}}$ gave a good fit to the line profile and so we also begin with these values.

\subsection{Geometry of the core and outflow}
Fig.~\ref{outflow_schematic} illustrates the geometry of the dynamical model we use here for the core and outflow. This geometric model is similar to that described in \citet{rawlings.et.al04}. The outflow shape is approximated geometrically using a tanh function to mimic the recollimation of the outflow a short distance from the central source. This gives the base of the outflow an hourglass-like morphology. The molecular outflow is confined to a layer close to the surrounding core and itself encloses a hot low density jet (not traced with this molecule in these transitions). The velocity of the outflow increases towards the outflow axis, as illustrated by the different length arrows in Fig.~\ref{outflow_schematic}. We include a freeze-out zone in the model where the CO undergoes a substantial depletion from the gas phase. The exact value of the abundance of the remaining gas phase CO in the frozen out zone is not critical to the model since it has a negligible effect on the line profiles.

\subsection{3D Radiative transfer code modelling}
The observational literature and analysis above serve to roughly constrain some of the physical parameters needed to solve the radiative transfer problem. One constraint on the CO abundances is set by measurements of the interstellar abundance ratios, from \citet{schoier.et.al} these are
\begin{equation}
      \; 	\; \; \; \; 	\frac {\rm C^{18}O} {\rm C^{17}O} = 3.9; \\
      \; 	\; \; \; \;	\frac {\rm ^{12}CO} {\rm ^{13}CO} =77; \\
      \; 	\; \; \; \;     \frac {\rm ^{13}CO}{\rm C^{18}O} = 7.3;
  \end{equation}
\begin{figure}
\centering
\includegraphics[width = 115 pt]{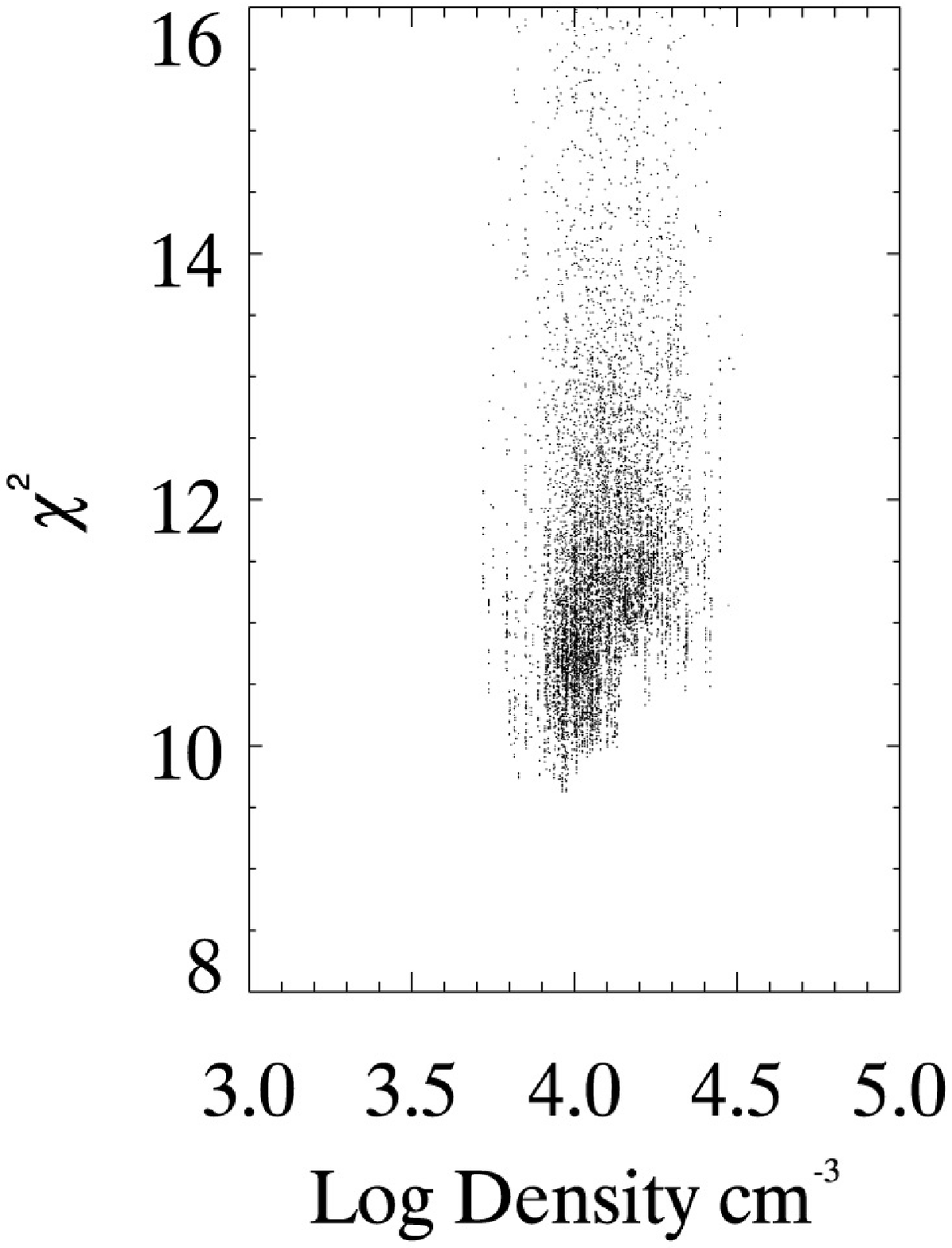}
\includegraphics[width = 115 pt]{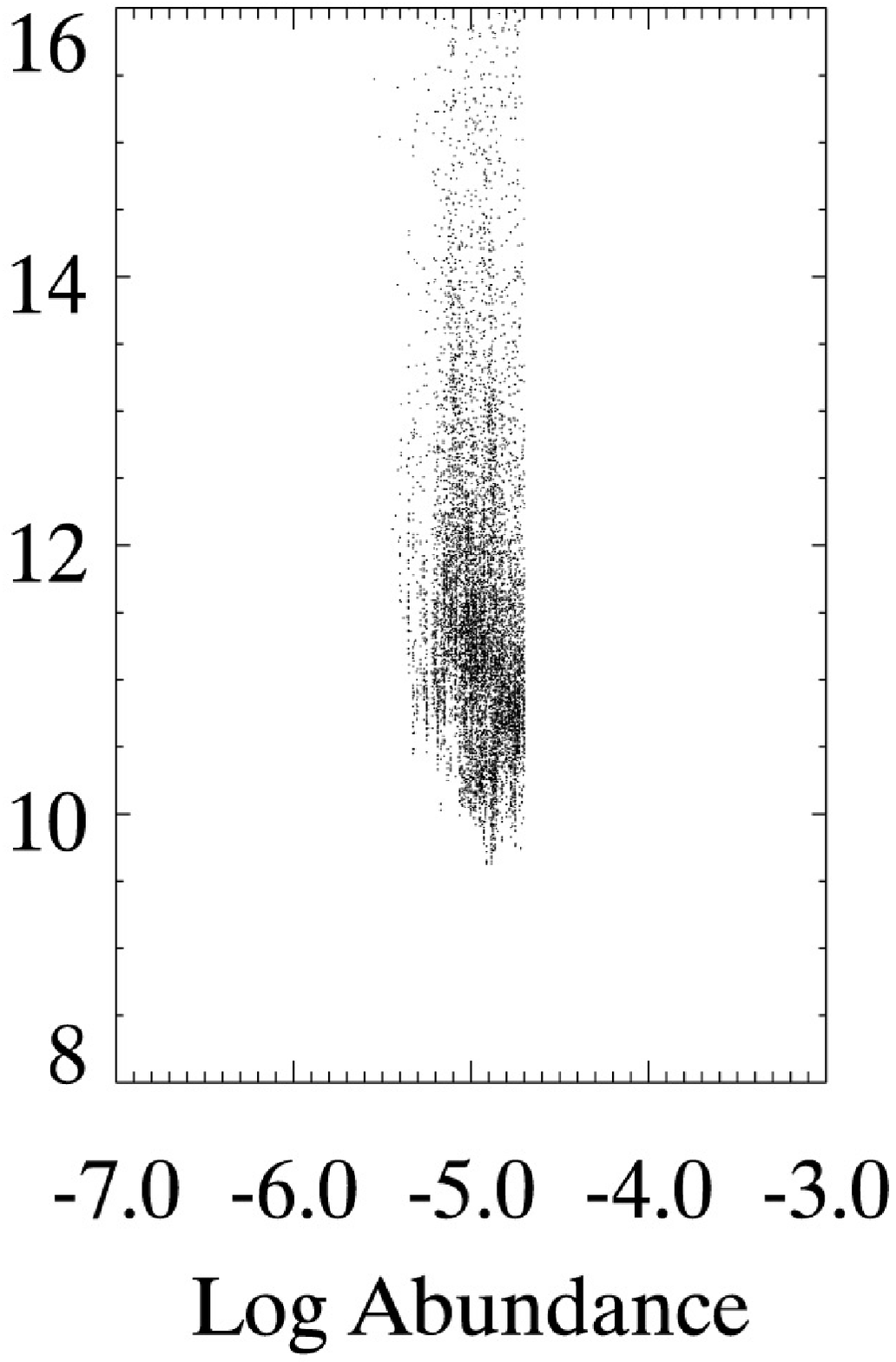}
\includegraphics[width = 150 pt]{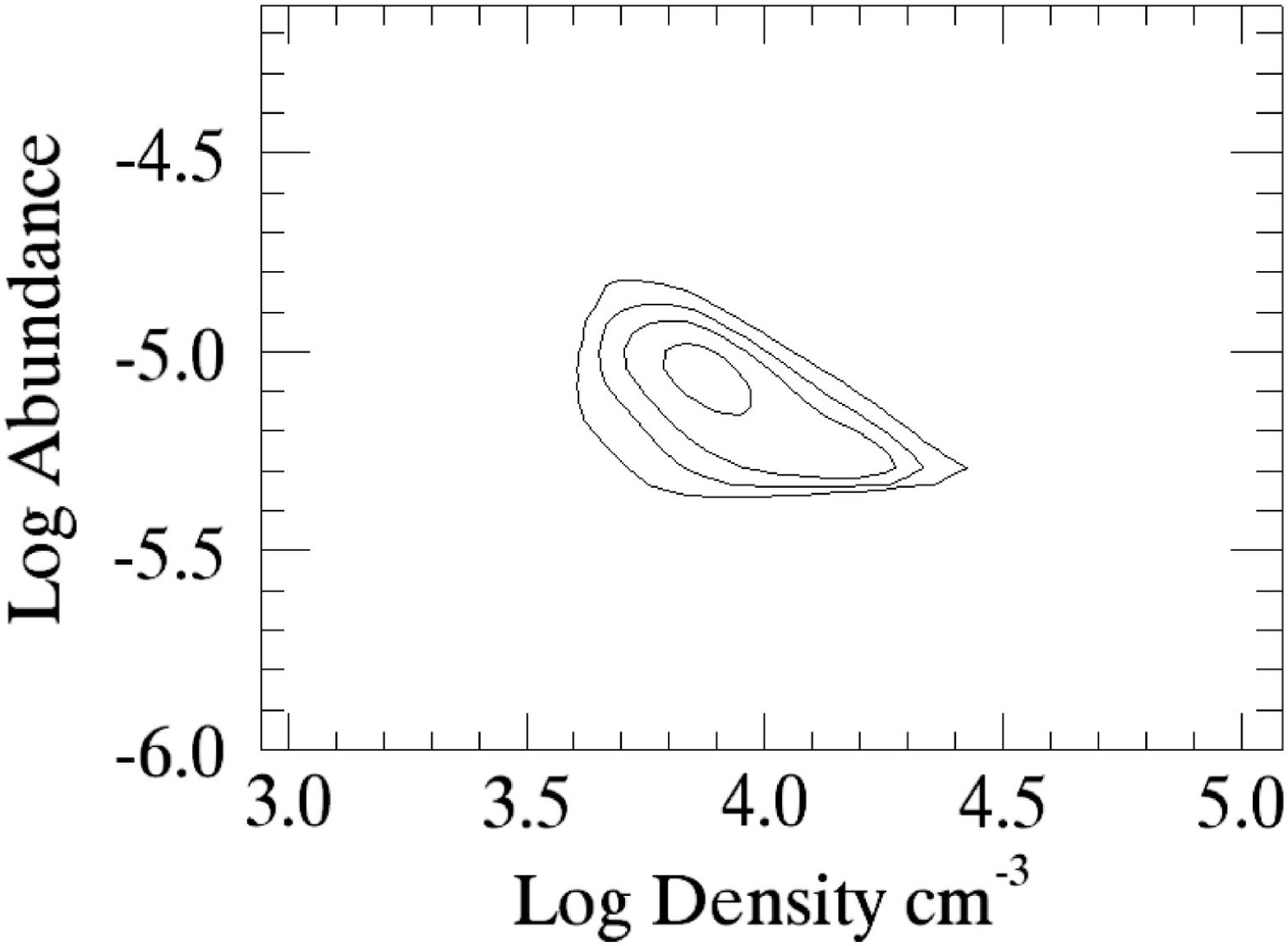}
\caption{The top two plots show the $\chi^{2}$ values versus the abundance and density repectively. The lowest $\chi^{2}$ values represents the input parameter used in our model. The third plot is the $\chi^{2}$ surface as a function of the two parameters. The contour values range form 10 - 14 in steps of 1 and a circular shape indicates no correlation between the paramters.}
\label{chi_squared_params}
\end{figure}
Figure~\ref{chi_squared_params} shows the $\chi^{2}$ procedure used to constrain the input parameters. The first two plots show how the $\chi^{2}$ of a model varies when one input parameter is held constant and all others are varied. The plots show the best $\chi^{2}$ and also how the variation in either parameter can cause a worse fit to the data. The bottom plot is the $\chi^{2}$ surface as s function of the abundance and density. A circular profile indicates no correlation exists amongst these two parameters which may be due to non-linear optical depth effects. We only show here the procedure carried out for the abundance and density in the inner boundary layer but this was carried out for for all the parameters listed in Table~\ref{parameters}.

Each model was used as an input to a 3D molecular line transport code (described in \citealt{keto,rawlings.et.al04,redman.et.al02b}) to produce synthetic line profiles for direct comparison with the observations. The integrated emission was also calculated in a given line of sight through the cloud. After a reasonable fit to the strength and width of the profiles was reached from varying the CO abundance the remaining quantities were varied over narrow ranges to improve the quality of the fits (for example, the turbulent width controls the thickness of the individual hyperfine lines seen in C$^{17}$O). It should be emphasized that the CO abundances were the dominant factor in the line profile formation, for example exact values of infall velocity had only a second order effect on the line strength, shape and width. 
\begin{table}
\centering
\caption{Best fit radiative transfer model parameters. The first four parameters were constrained by observations and fine tuned to improve the model fits. The abundances (defined as ${N_{\rm species}} / {N_{{\rm H}_{2}}}$) are our derived parameters. Entries marked with an asterisk are somewhat unconstrained by the model since there was little contribution to the line shape from these components. The density parameters for all three components refer to the peak value of a r$^{-2}$ power law density profile that has been truncated at a small radius.}
\begin{tabular}{@{}l l l l @{}}
\hline
Parameters& Envelope & Outer & Inner\\
& &	boundary & boundary \\
& &   layer			& layer\\
\hline
Temperature (K)&10&35&90\\
Peak Density (cm$^{-3}$)&$10^6$&$10^5$&$10^4$\\
Velocity (km s$^{-1}$)&$$-$0.6$&$1.9$&$3.5$\\
Turbulent &$0.1$&$0.3$&$1.4$\\
width (km s$^{-1}$)& & & \\
\hline
Abundance $^{12}$CO $(\times 10^{-7})$&$170$&$200$&$350$\\
Abundance $^{13}$CO $(\times 10^{-7})$&$2.5$&$6$&$80.0$\\
Abundance ${\rm C^{18}O}$ $(\times 10^{-7})$&$0.5$&$0.8^{*}$&$1.4^{*}$\\
Abundance ${\rm C^{17}O}$ $(\times 10^{-7})$&$0.1$&$0.33^{*}$&$0.56^{*}$\\
\hline
\end{tabular}
\label{parameters}
\end{table}

Table~\ref{parameters} lists the best fit parameters used in our final models. Examination of Table~\ref{parameters} shows that the boundary layers have systematically higher CO abundances than the envelope. The most obvious explanation for this is that the interaction of the jet with the envelope in the boundary layers (irrespective of the jet model) causes shock heating which in turn liberates molecular material that is frozen onto the dust grains. As already shown above the gas away from the boundary layer exhibits heavy CO depletion (Fig.~\ref{freezeout}). The temperature and CO abundance rises from 35K in the outer boundary layer adjacent to the envelope to 90K in the inner boundary layer towards the jet. 
The rise in modeled temperature and the changes in the CO abundances are consistent with our knowledge of CO sublimation temperatures. From Table~\ref{parameters} the abundance enhancement of $^{13}$CO going from the outer boundary layer to the much warmer inner boundary layer is nearly twice that of $^{12}$CO. The reason for this is likely to be complex involving chemical fractionation effects due to shock photons, selective desorption off the dust grains and grain surface chemistry.

\begin{figure}
\includegraphics[width = 260 pt]{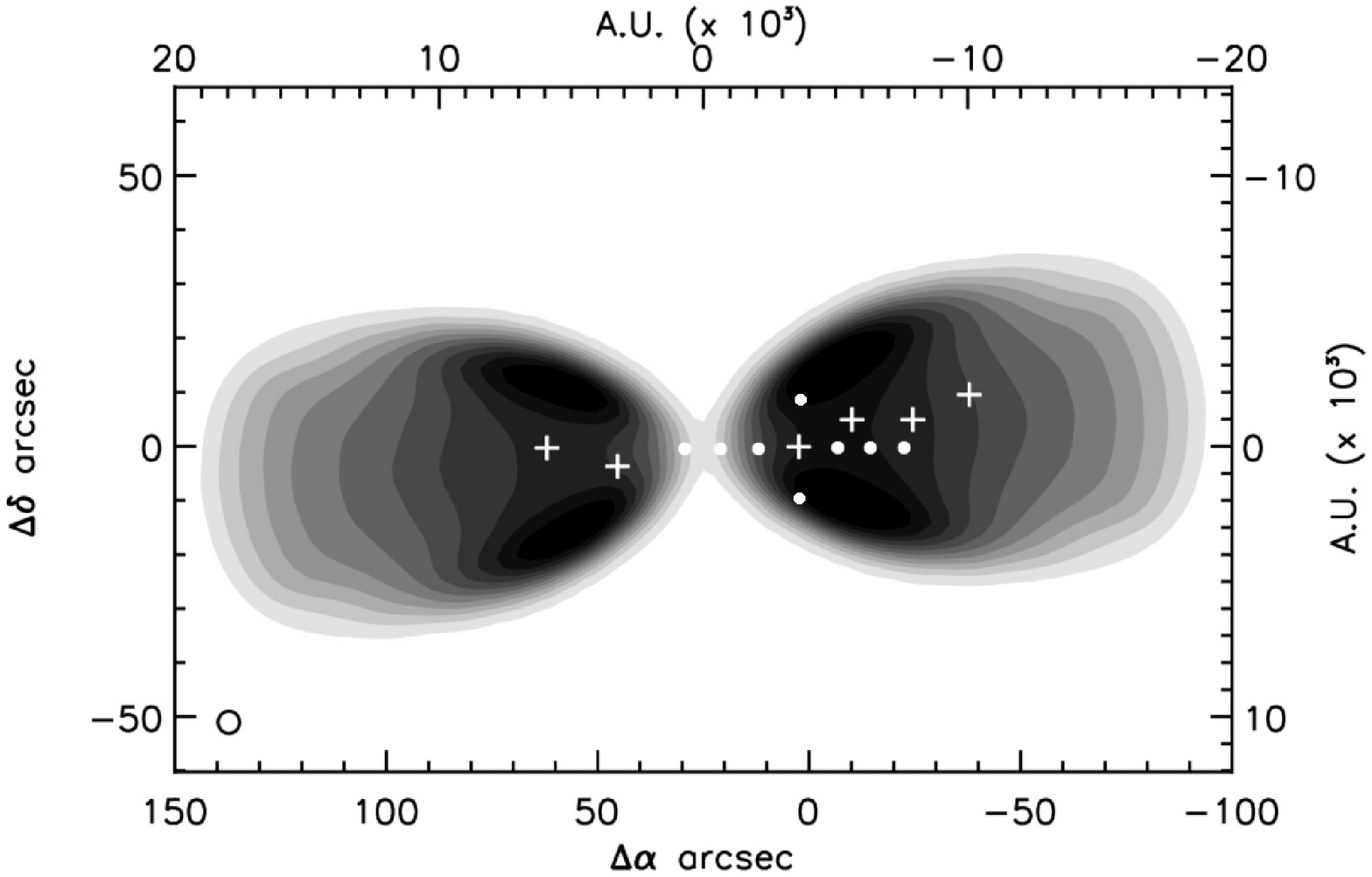}
\includegraphics[width = 260 pt]{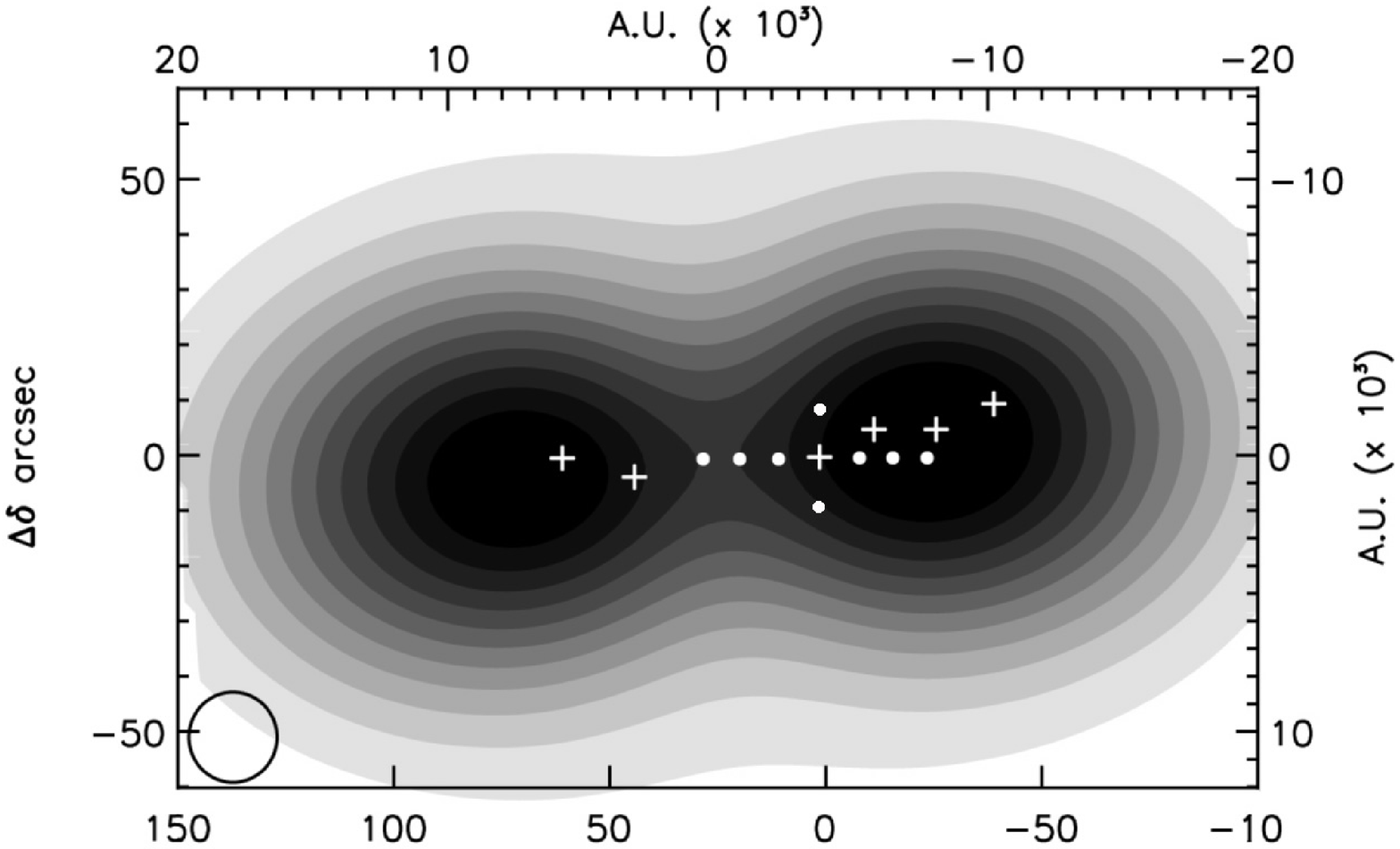}
\caption{Contour plot of $^{12}$CO model integrated emission from L483 highlighting the jet morphology. The top panel is modeled emission with a $\approx~1^{\prime\prime}$ beam, the bottom panel shows the emission convolved for a $\approx~15^{\prime\prime}$ beam. The emission includes both the red and blue-shifted components of the outflow. The + signs mark positions observed by J. Hatchell and the circles represent positioned observed in C$^{17}$O. Offsets are (RA, Dec) in arcseconds from RA = 18$^{h}$14$^{m}$50$^{s}$.6 Dec = -4$^{\circ}$40$^{\prime}$49$^{\prime\prime}$ (B1950).}
\label{filled_contour_posn_maps}
\end{figure}

We can compare our derived abundances in the envelope with those of \cite{jorgensen.et.al04}. They modeled L483 to find the following average abundances
\begin{equation}
\; \;\; \; \; \;\; \frac {\rm C^{18}O} {\rm H_2}  = 2.5  \times 10^{-8}; \\
\; \;\; \; \; \;\; \frac {\rm C^{17}O}{\rm H_2} = 7.8 \times 10^{-9}.
\label{jorgensen_ratios}
\end{equation}
These are broadly similar to, but lower than, our envelope values. Their modeled profiles underestimated the intensity with respect to observations. They concluded that the out-most part of the envelope may not have been accounted for in the models. A multi component model, such as that presented here, would be an alternative way to raise their derived abundances.

\begin{figure*}
\centering
\includegraphics[width = 500 pt]{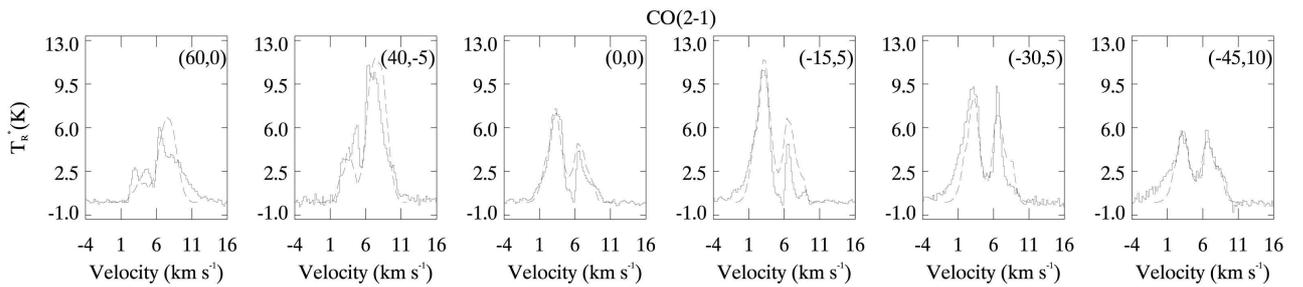}
\caption{$^{12}$CO (2 - 1) observed and modeled line profiles. The continuous line is the observed profile . The line profiles are from an approximately east-west string of positions across the molecular outflow. Their locations are marked with a '+' on Fig.~\ref{filled_contour_posn_maps} (data from \citealt{hatchell.et.al99})         }
\label{12co}
\end{figure*}

\subsection{Mass of the outflow}
Using Table~\ref{parameters} and our observations of the column density of H$_{2}$ (Fig.~\ref{freezeout}) we derived the column density of $^{13}$CO in the outflow. Integrating this over the spatial extent of the outflow \citep{margulis.85,parker.et.al91,coulson.et.04} then yields the mass of outflowing material. We include material affected by the jet plus boundary layer material from Table~\ref{parameters} to get the abundance and using an average value for N[H$_{2}$] we derive a $^{13}$CO outflow mass of $\approx 0.019$  solar masses over the extent of the outflow, we deduce the outflow energy to be $8.76\times10^{34}$ J and momentum $\sim 1.1\times10^{32}~{\rm kg~m~s^{-1}}$. These values are consistent with results obtained by integrating $^{13}$CO line wing emission, which is specifically caused by the outflowing component \citep{parker.et.al91}. L483 is on the transition between Class 0 and Class 1, meaning its core still resides within a substantial envelope. Estimating the gravitational binding energy of the cloud from our model parameters as $\sim 1.0\times10^{35}$ J, we can say the outflow kinetic energy is sufficient to play a role in disrupting the surrounding envelope.
 
\subsection{$^{12}$CO}
Fig.~\ref{filled_contour_posn_maps} shows  $^{12}$CO modeled integrated emission. The top map is the emission as would be seen with a high resolution ($\approx 1$$^{\prime\prime}$) instrument and on the bottom is the same emission as would be seen through a wider beam ($\approx 15$$^{\prime\prime}$) such as the JCMT. The outflow is at the same position angle as L483 and orientated with the right hand side lobe pointing 30$^{\circ}$ towards the observer as in Fig.~\ref{outflow_schematic}. The overall morphology shows a reasonable agreement with that observed in fig. A.2. of \citet{tafalla.et.al00}.

In Fig.~\ref{12co} we show the observed line with a continuous line and the overlayed modelled profile with a broken line. $^{12}$CO has contributions to its profile from all components, envelope, outer and inner boundary layer. Specifically each profile has two peaks, one larger than the other. The large peaks on the blue and red-shifted sides coincide with the location of maximum outflow emission and were used to constrain the inner boundary layer abundance and velocity. As seen in Fig.~\ref{filled_contour_posn_maps} the position of the $^{12}$CO profiles coincide with the peak emission of the outflow. We found only a high velocity component with significant abundance enhancement could account for the peak. The parameters for the inner boundary layer are those that model the contribution of the outflow to the line profile. Alongside the outflow peak is a smaller maximum which is closely delineated by parameters from the envelope and outer boundary layer regions. Such a big absorption dip indicates massive self absorption occurring in this species. It should be noted the envelope will have a small input to either side of the self absorption dip but the inner boundary layer component will totally dominate the emission and line width in the modelled profile. {${\rm ^{12}CO}~J=4\rightarrow 3$} data was also obtained for the same pointing positions as {${\rm ^{12}CO}~J=2\rightarrow 1$} and were well fit by our model but it is not shown here for clarity and conciseness.       

\subsection{$^{13}$CO}
\begin{figure}
\includegraphics[width = 250 pt]{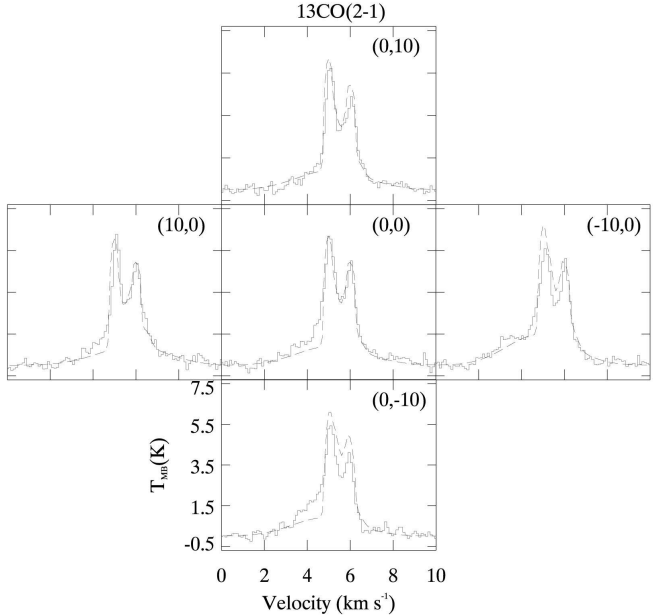}
\caption{$^{13}$CO (2 - 1) observed and modeled line profiles. The continuous line is the observed profile. The offset between cells is 10" and the middle panel is at 0" offset. }
\label{13co}
\end{figure}
The $^{13}$CO modeled profile closely matches the self absorption dip seen in the observed profile, Fig.~\ref{13co}. In optically thick species, after the onset of self-absorption, the intensity starts to become relatively insensitive to the overall molecular abundance \citep*{ward-thompson&buckley01,jorgensen}. This was the case for our $^{13}$CO models too. The blue asymmetric profile and the absorption dip could only be reproduced from the abundance and velocity parameters of the envelope. However, to account for the broadening in the blue-shifted side of the observed profiles (all five positions are on the blue-shifted side of the central source) we required a contribution from the fast flowing, warm inner boundary layer component. This contribution provides an extra constraint on the boundary layer abundance. 

\subsection{C$^{18}$O}
\begin{figure}
\includegraphics[width = 250pt]{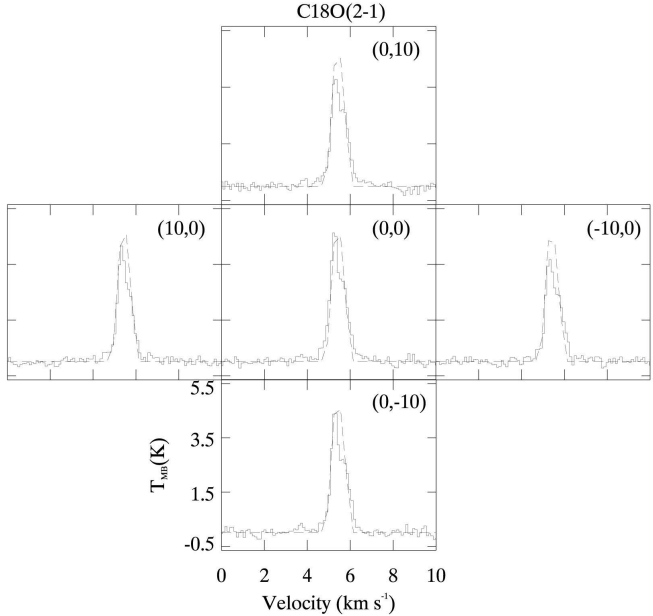}
\caption{C$^{18}$O (2 - 1) observed and modeled profiles. The continuous line is the observed profile. The offset between cells is 10".}
\label{c18o}
\end{figure}

The C$^{18}$O modeled profile in Fig.~\ref{c18o} exhibits a stronger blue than red side though there is no strong self-absorption dip in between. We cannot see any self absorption in C$^{18}$O because its abundance is not high enough to trigger this optical depth effect. It may also be that there is a small contribution on the blue side of the peak from the outflow since the five pointing positions are all on the blue-shifted side of the outflow. The peak of the intensity lies at the same position as the minimum of $^{13}$CO's self absorption dip. The line is not too optically thick and we found the abundance in the envelope was the main parameter controlling the shape of the modelled profile. Contributions from the inner or outer boundary layer have very little to no effect. This reaffirms the role of C$^{18}$O as a good tracer of static components residing deep inside the envelope.

\subsection{C$^{17}$O}
\begin{figure*}
\centering
\includegraphics[width = 500 pt]{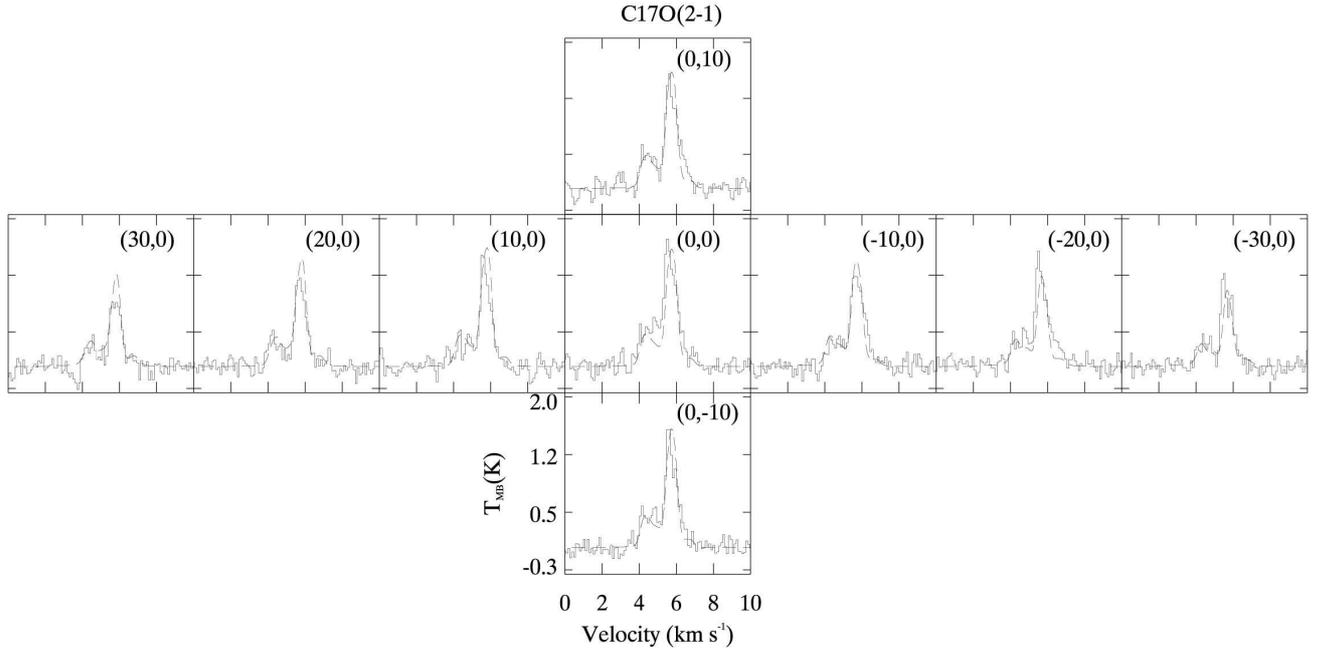}
\caption{C$^{17}$O (2 - 1) observed and modeled line profiles. The continuous line is the observed profile. The offset between cells is 10" and the middle panel is at 0" offset. Their locations are marked with a filled circle on Fig.~\ref{filled_contour_posn_maps}}
\label{c17o}
\end{figure*}

In Fig.~\ref{c17o} we observe the hyperfine structure induced by the $^{17}$O nucleus. The relative intensity of two hyperfine lines allow us to determine the optical depth in the cloud making it a very useful species to observe with. The hyperfine structure of C$^{17}$O was included in the R.T. code by using the line strengths and relative intensities from \citet*{ladd.et.al98} to construct the basic line shape that replaces the gaussians used for lines without hyperfine structure. The hyperfine components are discernible though there is some mild blending due to infall motion in the core. The boundary layer parameters have a negligible effect on the modeled line profiles with the turbulent velocity far more important.

\section{Discussion and conclusions}
We modelled L483 in four isotopomers of CO in order of decreasing abundance $^{12}$CO, $^{13}$CO, C$^{18}$O and C$^{17}$O. Each species delineates a different region and therefore we get a clearer picture of cloud dynamics.

C$^{18}$O and C$^{17}$O are optically thin and are important tracers of the denser regions of star forming clouds. Optically thick species such as $^{12}$CO and $^{13}$CO are more abundant and more susceptible to the jet motion. They trace regions farther out from the centre of the cloud and have line profiles showing a large line width consistent with material exposed to a high velocity jet. Using a radiative transfer code and a dynamical model for L483 we were able to self-consistently calculate for the first time the abundance of the CO isotopes in the different regions of such a cloud.

Our principal finding  is that the CO line profiles in L483 are well fitted with a self-consistent envelope plus boundary layer model and that the CO abundances increase substantially in this boundary layer. The most likely reason for this is that molecular ices on dust grains are heated and released back into the gas phase in the boundary layer. A constant abundance model was found to overestimate the abundance towards the centre of the cloud and only freeze-out of material towards the centre was able to produce modeled profiles consistent with observations. Our tanh geometry is chosen because it matches the observed morphology seen in other protostellar outflows \citep{tafalla.et.al00}.  Other geometries are possible for the outflow e.g. conical, cylindrical outflow but it is unlikely that it would have a substantial effect on our results because such a detailed treatment of the boundary layer components is difficult to achieve until there is sub-arcsecond resolution resolution. 

We emphasize that our results provide an abundance enhancement measurement rather than proving an exact mechanism by which the CO is enhanced, e.g. chemical reactions or dissociation. A more detailed treatment would involve a full dark cloud and gas-grain chemistry whilst accounting for localized shock heating. The most enhanced species in our study, by a factor of $\sim 30$, is the $^{13}$CO material. The exothermic reaction leading to the creation of $^{13}$CO \citep{duley_williams} is shown below
\begin{equation}
 \; \; \; \; \; ^{13}{\rm C}^{+} + ^{12}{\rm CO} \rightleftharpoons ^{12}{\rm C}^{+} + ^{13}{\rm CO} + \Delta{\rm E}
\label{13co_creation}
\end{equation}
where the zero-point energy difference $\Delta{\rm E}$ is equivalent to a temperature $\Delta{\rm E}/{\rm k}$ of 35~K. This mechanism may be the source for the enhanced abundance observed from our modeling.

The enhanced abundance seen in the boundary layer effect may also be detectable in other molecules. \citet{park.et.al00} used interferometric observations of HCO$^{+}$ and observed anti-infall profiles close to the centre of the cloud. They concluded the HCO$^{+}$ was tracing the outlying regions of the outflow, i.e. a region between the envelope and the jet. The reason the HCO$^{+}$ emission is predominately seen here rather than in the more extensive envelope is also likely due to an enhancement of HCO$^{+}$ caused by the shock-heated release of icy grain mantles followed by chemical reaction. CO and H$_2$O are liberated into the gas phase and the shock-induced radiation field then can photodissociate CO to C$^+$. This then reacts with the H$_2$O to form HCO$^{+}$. Such a model was successfully used to explain the enhancement of HCO$^{+}$ commonly seen at the bases of molecular outflows  \citep{rawlings.et.al00,rawlings.et.al04}. The results in this paper demonstrate that a combination of datasets with several lines and transitions coupled with a 3D molecular line transport code is a powerful way to determine the properties of dense star forming cores.
\section*{Acknowledgments} 
We thank the anonymous referee for constructive  comments that helped to significantly improve the paper. We also thank Jennifer Hatchell for kindly supplying us with reduced {${\rm ^{12}CO}$} data of L483. PBC received financial support from the Cosmo-Grid project, funded by the Program for Research in Third Level Institutions under the National Development Plan and with assistance from the European Regional Development Fund. MPR was supported by the UK PPARC, and through an IRCSET Ireland fellowship during the early stages of this work. We thank the staff of the JCMT and visiting observers for obtaining the observations. The JCMT is operated by The Joint Astronomy Centre on behalf of the Science and Technology Facilities Council of the United Kingdom, the Netherlands Organization for Scientific Research, and the National Research Council of Canada.
\bibliographystyle{mn2e} \bibliography{l483}

\begin{thebibliography}{}

\bibitem[\protect\citeauthoryear{{Bacmann}, {Lefloch}, {Ceccarelli}, {Castets},
  {Steinacker} \& {Loinard}}{{Bacmann} et~al.}{2002}]{bacmann.et.al02}
{Bacmann} A.,  {Lefloch} B.,  {Ceccarelli} C.,  {Castets} A.,  {Steinacker} J.,
     {Loinard} L.,  2002, A\&A, 389, L6

\bibitem[\protect\citeauthoryear{{Beltr{\' a}n}, {Estalella}, {Anglada},
  {Rodr{\'{\i}}guez} \& {Torrelles}}{{Beltr{\' a}n}
  et~al.}{2001}]{beltran.et.al01}
{Beltr{\' a}n} M.~T.,  {Estalella} R.,  {Anglada} G.,  {Rodr{\'{\i}}guez}
  L.~F.,    {Torrelles} J.~M.,  2001, AJ, 121, 1556

\bibitem[\protect\citeauthoryear{{Buckle}, {Hatchell} \& {Fuller}}{{Buckle}
  et~al.}{1999}]{buckle.et.al99}
{Buckle} J.~V.,  {Hatchell} J.,    {Fuller} G.~A.,  1999, A\&A, 348, 584

\bibitem[\protect\citeauthoryear{{Coulson}, {Dent} \& {Greaves}}{{Coulson}
  et~al.}{2004}]{coulson.et.04}
{Coulson} I.~M.,  {Dent} W.~R.~F.,    {Greaves} J.~S.,  2004, MNRAS, 348, 39

\bibitem[\protect\citeauthoryear{Duley \& Williams}{Duley \&
  Williams}{1984}]{duley_williams}
Duley W.~W.,  Williams D.~A.,  1984, Insterstellar Chemistry.
Academic Press

\bibitem[\protect\citeauthoryear{{Fuller}, {Lada}, {Masson} \&
  {Myers}}{{Fuller} et~al.}{1995}]{fuller.et.al95}
{Fuller} G.~A.,  {Lada} E.~A.,  {Masson} C.~R.,    {Myers} P.~C.,  1995, ApJ,
  453, 754

\bibitem[\protect\citeauthoryear{{Fuller} \& {Myers}}{{Fuller} \&
  {Myers}}{1992}]{fuller.et.92}
{Fuller} G.~A.,  {Myers} P.~C.,  1992, ApJ, 384, 523

\bibitem[\protect\citeauthoryear{{Fuller} \& {Wootten}}{{Fuller} \&
  {Wootten}}{2000}]{fuller&wootten00}
{Fuller} G.~A.,  {Wootten} A.,  2000, ApJ, 534, 854

\bibitem[\protect\citeauthoryear{{Gibb} \& {Little}}{{Gibb} \&
  {Little}}{1998}]{gibb}
{Gibb} A.~G.,  {Little} L.~T.,  1998, MNRAS, 295, 299

\bibitem[\protect\citeauthoryear{{Goldsmith}, {Li} \& {Kr{\v c}o}}{{Goldsmith}
  et~al.}{2007}]{goldsmith_07}
{Goldsmith} P.~F.,  {Li} D.,    {Kr{\v c}o} M.,  2007, ApJ, 654, 273

\bibitem[\protect\citeauthoryear{{Hatchell}, {Fuller} \& {Ladd}}{{Hatchell}
  et~al.}{1999}]{hatchell.et.al99}
{Hatchell} J.,  {Fuller} G.~A.,    {Ladd} E.~F.,  1999, A\&A, 344, 687

\bibitem[\protect\citeauthoryear{{J{\o}rgensen}, {Hogerheijde}, {Blake}, {van
  Dishoeck}, {Mundy} \& {Sch{\"o}ier}}{{J{\o}rgensen}
  et~al.}{2004}]{jorgensen.et.al04}
{J{\o}rgensen} J.~K.,  {Hogerheijde} M.~R.,  {Blake} G.~A.,  {van Dishoeck}
  E.~F.,  {Mundy} L.~G.,    {Sch{\"o}ier} F.~L.,  2004, A\&A, 415, 1021

\bibitem[\protect\citeauthoryear{{J{\o}rgensen}, {Schoier} \& {van
  Dishoeck}}{{J{\o}rgensen} et~al.}{2002}]{jorgensen}
{J{\o}rgensen} J.~K.,  {Schoier} F.~L.,    {van Dishoeck} E.~F.,  2002, A\&A,
  389, 908

\bibitem[\protect\citeauthoryear{{Keto}, {Rybicki}, {Bergin} \& {Plume}}{{Keto}
  et~al.}{2004}]{keto}
{Keto} E.,  {Rybicki} G.~B.,  {Bergin} E.~A.,    {Plume} R.,  2004, ApJ, 613,
  355

\bibitem[\protect\citeauthoryear{{Ladd}, {Adams} \& {Fuller}}{{Ladd}
  et~al.}{1991}]{ladd.et.91}
{Ladd} E.~F.,  {Adams} F.~C.,    {Fuller} G.~A.,  1991, ApJ, 382, 555

\bibitem[\protect\citeauthoryear{Ladd, Fuller \& Deane}{Ladd
  et~al.}{1998}]{ladd.et.al98}
Ladd E.~F.,  Fuller G.~A.,    Deane J.~R.,  1998, ApJ, 495, 871

\bibitem[\protect\citeauthoryear{{Lee}, {Bergin} \& {Evans}}{{Lee}
  et~al.}{2004}]{lee_04}
{Lee} J.-E.,  {Bergin} E.~A.,    {Evans} N.~J.,  2004, ApJ, 617, 360

\bibitem[\protect\citeauthoryear{{Margulis} \& {Lada}}{{Margulis} \&
  {Lada}}{1985}]{margulis.85}
{Margulis} M.,  {Lada} C.~J.,  1985, ApJ, 299, 925

\bibitem[\protect\citeauthoryear{{Myers}, {Mardones}, {Tafalla}, {Williams} \&
  {Wilner}}{{Myers} et~al.}{1996}]{myers.96}
{Myers} P.~C.,  {Mardones} D.,  {Tafalla} M.,  {Williams} J.~P.,    {Wilner}
  D.~J.,  1996, ApJ, 465, 133

\bibitem[\protect\citeauthoryear{{Nisini}, {Codella}, {Giannini}, {Santiago
  Garcia}, {Richer}, {Bachiller} \& {Tafalla}}{{Nisini}
  et~al.}{2007}]{nisini_07}
{Nisini} B.,  {Codella} C.,  {Giannini} T.,  {Santiago Garcia} J.,  {Richer}
  J.~S.,  {Bachiller} R.,    {Tafalla} M.,  2007, A\&A, 462, 163

\bibitem[\protect\citeauthoryear{{Park}, {Kim} \& {Minh}}{{Park}
  et~al.}{1999}]{park.et.al99}
{Park} Y.-S.,  {Kim} J.,    {Minh} Y.~C.,  1999, ApJ, 520, 223

\bibitem[\protect\citeauthoryear{{Park}, {Panis}, {Ohashi}, {Choi} \&
  {Minh}}{{Park} et~al.}{2000}]{park.et.al00}
{Park} Y.-S.,  {Panis} J.-F.,  {Ohashi} N.,  {Choi} M.,    {Minh} Y.~C.,  2000,
  ApJ, 542, 344

\bibitem[\protect\citeauthoryear{{Parker}, {Padman} \& {Scott}}{{Parker}
  et~al.}{1991}]{parker.et.al91}
{Parker} N.~D.,  {Padman} R.,    {Scott} P.~F.,  1991, MNRAS, 252, 442

\bibitem[\protect\citeauthoryear{Rawlings, Redman, Keto \& Williams}{Rawlings
  et~al.}{2004}]{rawlings.et.al04}
Rawlings J. M.~C.,  Redman M.~P.,  Keto E.,    Williams D.~A.,  2004, MNRAS,
  351, 1054

\bibitem[\protect\citeauthoryear{Rawlings, Taylor \& Williams}{Rawlings
  et~al.}{2000}]{rawlings.et.al00}
Rawlings J. M.~C.,  Taylor S.~D.,    Williams D.~A.,  2000, MNRAS, 313, 461

\bibitem[\protect\citeauthoryear{Redman, Rawlings, Nutter, Ward-Thompson \&
  Williams}{Redman et~al.}{2002}]{redman.et.al02b}
Redman M.~P.,  Rawlings J. M.~C.,  Nutter D.~J.,  Ward-Thompson D.,    Williams
  D.~A.,  2002, MNRAS, 337, L17

\bibitem[\protect\citeauthoryear{{Sandford} \& {Allamandola}}{{Sandford} \&
  {Allamandola}}{1993}]{sandford}
{Sandford} S.~A.,  {Allamandola} L.~J.,  1993, ApJ, 417, 815

\bibitem[\protect\citeauthoryear{{Schoier}, {J{\o}rgensen}, {Dishoeck} \&
  {Blake}}{{Schoier} et~al.}{2002}]{schoier.et.al}
{Schoier} F.~L.,  {J{\o}rgensen} J.~K.,  {Dishoeck} E.~F.,    {Blake} G.~A.,
  2002, A\&A, 390, 1001

\bibitem[\protect\citeauthoryear{Shirley, Evans, Rawlings \& Gregersen}{Shirley
  et~al.}{2000}]{shirley.et.al00}
Shirley Y.~L.,  Evans N.~J.,  Rawlings J. M.~C.,    Gregersen E.,  2000, ApJS,
  131, 249

\bibitem[\protect\citeauthoryear{{Tafalla}, {Myers}, {Caselli}, {Walmsley} \&
  {Comito}}{{Tafalla} et~al.}{2002}]{tafalla.et.al02}
{Tafalla} M.,  {Myers} P.~C.,  {Caselli} P.,  {Walmsley} C.~M.,    {Comito} C.,
   2002, ApJ, 569, 815

\bibitem[\protect\citeauthoryear{{Tafalla}, {Myers}, {Mardones} \&
  {Bachiller}}{{Tafalla} et~al.}{2000}]{tafalla.et.al00}
{Tafalla} M.,  {Myers} P.~C.,  {Mardones} D.,    {Bachiller} R.,  2000, A\&A,
  359, 967

\bibitem[\protect\citeauthoryear{{van Dishoeck}, {Blake}, {Jansen} \&
  {Groesbeck}}{{van Dishoeck} et~al.}{1995}]{van_dishoeck_95}
{van Dishoeck} E.~F.,  {Blake} G.~A.,  {Jansen} D.~J.,    {Groesbeck} T.~D.,
  1995, ApJ, 447, 760

\bibitem[\protect\citeauthoryear{Ward-Thompson \& Buckley}{Ward-Thompson \&
  Buckley}{2001}]{ward-thompson&buckley01}
Ward-Thompson D.,  Buckley H.~D.,  2001, MNRAS, 327, 955

\end{thebibliography}
\end{document}